\documentclass[fleqn,twoside]{article}
 \usepackage{espcrc2}
\readRCS
$Id: espcrc2.tex,v 1.2 2004/02/24 11:22:11 spepping Exp $
\usepackage{graphicx}
\usepackage[figuresright]{rotating}

\newcommand{\AmS}{{\protect\the\textfont2
  A\kern-.1667em\lower.5ex\hbox{M}\kern-.125emS}}

\def\p{I\!\!P}

\def\beq{\begin{equation}}
\def\eeq{\end{equation}}
\def\bea{\begin{eqnarray}}
\def\eea{\end{eqnarray}}
\title{High-energy photon collisions at the LHC - dream or reality?}

\author{M.\ Klasen
 \address{Laboratoire de Physique Subatomique et de Cosmologie,
 Universit\'e Joseph Fourier / CNRS-IN2P3 / INPG, 53 Avenue des Martyrs,
 F-38026 Grenoble, France}%
 \thanks{Research supported by the French Ministry for Higher Education and
 Research, CNRS-IN2P3 and ANR.}}

\runtitle{High-energy photon collisions at the LHC - dream or reality?}
\runauthor{M.\ Klasen}

\begin{document}

\begin{abstract}
 We discuss the potential of high-energy photon collisions
 at the LHC for improving our understanding of QCD and studying
 the physics beyond the Standard Model. After reviewing briefly the
 legacy of past photoproduction experiments at LEP and HERA,
 we examine the gold-plated channels proposed for a photon collider
 at the ILC for their potential in a hadron collider environment. We
 stress that initial-state photon interactions have indeed been
 observed at RHIC and at the Tevatron. Three promising channels at
 the LHC are then presented in some detail: exclusive vector-meson
 production, measurements of possibly anomalous electroweak
 gauge-boson or top-quark couplings, and slepton production.
\vspace{1pc}
\end{abstract}

\maketitle

\vspace*{-100mm}
\noindent LPSC 08-054
\vspace*{80mm}

\section{Introduction}

Until quite recently, high-energy photon interactions have been the
exclusive domain of lepton accelerators. The emission of photons by
electron beams has been known for many years to be not only a
nuisance to accelerator physicists (although many interesting results
have been recently obtained using initial-state radiation at
$B$-factories), but also a useful tool for
various domains of science, including biophysics and material
science, but also particle physics. It was hoped that an
International Linear Collider (ILC) might be built soon and include a
laser-backscattering facility, turning it into a photon
collider with up to 80\% of the center-of-mass energy of the parent
ILC.

The recent funding cuts in the United Kingdom and the United States
represent unfortunately serious set-backs for the prospects of
studying photon interactions at lepton colliders. The search for
short- or medium-term alternatives has thus become imperative. With
the observation of photoproduction events at existing hadron
colliders such as RHIC or the Tevatron and the imminent start-up of
the LHC, the latter has emerged as a serious candidate for studying
high-energy photon collisions in the near future.

In this review, we motivate these studies beyond the famous quote
from Genesis 1, 3-4, where {\em God said: `Let there be light,' and
there was light. God saw that the light was good, and he separated
the light from the darkness.} We first discuss the legacy of past
photoproduction experiments at LEP and HERA and the gold-plated
channels proposed for a photon collider at the ILC. We then present
the emerging experimental results from RHIC and the Tevatron
and three promising channels at the LHC: exclusive vector-meson
production, measurements of possibly anomalous electroweak
gauge-boson or top-quark couplings, and slepton production. We
close with an outlook on other channels that might be interesting
to study further in the future.

\section{Photon physics at lepton colliders}

At circular lepton accelerators such as LEP and the electron-ring
at HERA, spacelike, almost real bremsstrahlung photons are
exchanged during the hard collision. At a future ILC, large
particle bunch densities are needed to reach high luminosities.
Then additional beamstrahlung photons will be created before the
hard interaction by the coherent action of the electromagnetic
field of one bunch on the opposite one. If the electron beams are
collided with additional high-energy laser beams, real photons can
be produced through Compton scattering. Thus three different
mechanisms contribute to photon scattering at high-energy lepton
colliders: bremsstrahlung, beamstrahlung, and laser backscattering.

\subsection{LEP}

The LEP collider was operated at CERN in two phases, i.e.\ in the
years 1989 to 1996 at a center-of-mass energy of 91 GeV and from
1998 to 2000 at center-of-mass energies between 161 and 209 GeV.
The photoproduction events allowed for a large variety of QCD
studies. They were divided into three categories: those with no,
one, or two identified electrons (positrons).

Untagged events were e.g.\ exploited by the L3 and OPAL
collaborations for measuring the total photon-photon cross section
with the result that both collaborations found consistently
evidence for soft-pomeron and reggeon, but no hard-pomeron
exchanges \cite{przbycien}. They disagreed, however, on the
inclusive-jet production cross section, where only OPAL found
agreement with the next-to-leading order (NLO) QCD prediction
\cite{kkk}. Similar agreement was found by OPAL for the dijet
cross section, whose direct, single-resolved, and double-resolved
components could be separated by reconstructing the observed photon
momentum fractions $x_{\gamma}^{\pm}$ from the transverse momenta
$p_T$ and rapidities $\eta$ of the final jets. Both L3 and OPAL
measured inclusive hadron photoproduction, finding (in the case of
L3 considerably) harder $p_T$-spectra than observed in
photon-proton collisions at HERA. This could, at least partly in
the case of L3, be attributed to the direct interactions of the two
photons.

In single-tagged events, where the virtuality $Q$ of one of the
photons provides a hard scale, inclusive cross section measurements
allowed to improve considerably our knowledge about the QED and
hadronic structure of the photon. The light, but also heavy (charm)
quark densities $f_{q/\gamma}$ became directly accessible down to
relatively low values of $x_\gamma\simeq0.01$, and QCD evolution
could be tested up to $Q\simeq20$ GeV. In this way, the gluon
density could at least be indirectly constrained and even the
strong coupling constant $\alpha_s$ could be measured with
competitive accuracy \cite{aks}. Recently, parton density function
(PDF) analyses for the proton have been performed at the
next-to-next-leading order (NNLO) level. In the future, this
should, of course, also be done for photons, where the direct
contribution plays a particularly important role \cite{hejbal}.

The direct contribution rises, of course, in the transition region
from real to virtual photons, which has been studied at LEP through
double-tagged events, whereas the genuine hadronic (vector-meson
dominance) contribution diminishes. The QED and hadronic structure
functions of virtual photons, or equivalently electrons, have been
extracted, and also the total $\gamma^*\gamma^*$ cross section has
been measured. For a more complete experimental review see 
\cite{nisius}.

\subsection{HERA}

At HERA, where electrons of energy 27.5 GeV were collided from 1992
to 2000 with protons of energy 820 GeV and from 2003 to 2007 with
protons of 920 GeV, photoproduction events were abundant and
allowed for QCD studies in at least four different respects
\cite{schoerner}. We concentrate in the following on dijet
production, but similar discussions also apply to light- and
heavy-flavor and prompt-photon production.

First, the distribution in the center-of-mass scattering angle
permitted to distinguish regions, where the dijet cross section was
dominated by spin-1/2 quark exchanges, from those, where it was
dominated by spin-1 gluon exchanges. Both distributions were very
different from the pure phase space distribution and provided
evidence of the underlying QCD dynamics.

Second, a separation of low- (less than 0.1) and high-$x_p$
(more than 0.1) contributions allowed not only to confirm the
shape of the quark distributions in the photon determined at LEP
(see above), but also to learn more about the photon's gluon
distribution than could be achieved in $e^+e^-$ collisions.

Third, restricting the measurements to high-$x_\gamma$
contributions (more than 0.8) permitted to obtain information
on the proton's gluon distribution that was complementary to
the one obtained in deep-inelastic scattering (DIS).

Fourth, the production of forward jets in the transition region
from real to virtual photons offered the interesting possibility
to compare the predictions made by the BFKL ($x$) and DGLAP ($Q^2$)
evolution equations, respectively.

The question whether diffractive dijet photoproduction can be
described by factorizing the hadronic cross section into universal
pomeron fluxes, PDFs and a perturbative partonic cross section or
is rather subject to initial-state rescattering is still under
discussion \cite{bruni}. At least the H1 data seem to indicate a
global suppression (or rapidity-gap survival probability) of about
0.5 of the NLO cross section \cite{kkdiff}, whereas theoretical
predictions based on a two-channel eikonal model predict a
suppression by about a factor of three for the resolved-photon
contribution \cite{kaidalov}.

Exclusive production of vector mesons or deeply virtual Compton
scattering may actually be easier to understand in the sense that
consistent values of the $t$-slope parameter or interaction size
$b$ of the pomeron can be extracted \cite{bunyatyan} or
generalized PDFs may be extracted \cite{pire}.

While the final HERA data has reached a quite good experimental
accuracy, in particular thanks to the high-luminosity running in
the second phase, the theoretical predictions still suffer from a
variety of uncertainties, even though virtually all two-to-two
photoproduction processes have now been calculated and
cross-checked at NLO \cite{chyla}.

The choice of renormalization/factorization schemes e.g.\ is of
particular importance in photoproduction due to the direct-photon
initial-state singularity, which can be more effectively resummed
in the DIS$_\gamma$ than in the $\overline{\rm MS}$ scheme. For
heavy-quark production, considerable progress has recently been
made by interpolating between the massive and zero-mass variable
flavor schemes through the general-mass variable flavor scheme.

Of equal importance in principle and even larger importance in
practice is the choice of renormalization/factorization scales,
which are often identified and varied by a factor of two about the
hardest scale in the scattering process, but could arguably be set
more efficiently by identifying the saddle point in a
two-dimensional scan. The scale uncertainties now being generally
larger than the statistical (and also systematic) experimental
uncertainties, it would certainly be desirable to be able to
compare the data to NNLO calculations,
but these are still to be completed. In the meantime, resummation
and NLO Monte Carlo (MC) programs offer higher precision than
NLO calculations, but attention in these fields has so far been
focused on LHC-relevant signal and background processes for new
physics. An experimental alternative would have been to move to
higher scales ($p_T$), but this was not always possible at HERA due
to the limited event rates. For a more complete review see
\cite{klasenreview}.

\subsection{ILC}

Whereas photon collisions in circular accelerators are limited by
the bremsstrahlung spectrum in energy and luminosity, the linear
beams at an ILC could be transformed through Compton backscattering
into high-energy and high-luminosity photon beams. Adjusting the
laser energy, polarization, distance from the interaction point and
crossing angle allows to further improve the performance of such a
photon collider \cite{telnov}.

Such a machine requires, if not a dedicated detector design, at
least some modifications for the proposed ILC detectors
\cite{gronberg}. In particular, more space and shielding has to be
provided for the beam pipe, affecting also the layout of the endcap
calorimeters. The desired performances are similar, with e.g.\ a 3\%
jet energy resolution, 5 $\mu$m vertex-tagging for $b$-quarks and
good hermiticity for supersymmetric (SUSY) events involving missing
(transverse) energy. Currently, three detector concepts are being
developed, one in the US with a silicon tracker (SiD) and two in
the EU (LDC) and Asia (GLD) with a time projection chamber, with
the latter two bound to merge soon into a common design (ILD).

The most promising physics case for a photon collider has always
been a precise Higgs-boson mass measurement (a precision of about 100 MeV in one
year of running), made possible through the reaction \cite{roeck}
\bea
 \gamma\gamma&\to&h~(H,A)~\to~b\bar{b}.
\eea
Other quantities that could thus be determined with high precision
include the Higgs partial widths to two photons, $W$- and
$Z$-bosons or top quarks. Should SUSY particles exist, pair
production of the heavier SM partners such as selectrons might not
be accessible with the limited ILC energy. In this case, the
associated production of a light neutralino with a heavier selectron
in photon-electron collisions
would represent another promising possibility allowing to search for
physics beyond the SM. Also, left- and right-handed anomalous $Wtb$
couplings might be determined with better accuracies than possible
certainly at the Tevatron, but also at the LHC and even in $e^+e^-$
collisions.

A more complete list of ``gold-plated'' channels at a photon
collider has been compiled in Vol.\ 6 of the TESLA TDR, shown here
as Tab.\ 1 \cite{teslatdr}.
%
%%% Begin  Table 1 %%%
\begin{table*}
 \begin{center}
 \caption{``Gold-plated'' channels at a photon collider and their
 potential at an ILC as well as potential
 difficulties in their adaption to the LHC and/or pertinent
 contribution in these proceedings.}
 \begin{tabular}{lll}
\hline
{\bf Channel} & {\bf ILC potential} & {\bf LHC potential} \\
\hline\hline
$\gamma\gamma\to h \to b\bar{b}$   & SM (or MSSM) Higgs, $M_h<160$ GeV &
 $\p\gg\gamma$, \\
$\gamma\gamma\to h \to WW(WW^*)$ & SM Higgs, 140 GeV $<M_h<190$ GeV &
 needs survival  \\
$\gamma\gamma\to h \to ZZ(ZZ^*)$ & SM Higgs, 180 GeV $<M_h<350$ GeV &
 probab.\ $S^2$ \cite{sarycheva,ovyn} \\
\hline
$\gamma\gamma\to H,A\to b\bar{b}$  &
 MSSM heavy Higgs, for intermediate $\tan\beta$ & similar to $h$ \\
$\gamma\gamma\to \tilde{f}\tilde{f}^*,\
\tilde{\chi}^+_i\tilde{\chi}^-_i,\ H^+H^-$ & Large cross sections,
possible observations of FCNC & \cite{gronberg,schul} \\ 
$\gamma\gamma\to S[\tilde{t}\tilde{t}^*]$ & 
$\tilde{t}\tilde{t}^*$ stoponium  & ? \\
$\gamma e^- \to \tilde{e}^- \tilde{\chi}_1^0$ & $M_{\tilde{e}^-} < 
0.9\times 2E_0 - M_{\tilde{\chi}_1^0}$ & ? \\
\hline
$\gamma\gamma\to W^+W^-$ & Anomalous $W$ interactions, extra dimensions & 
 \cite{manteuffel,pierzchala} \\
$\gamma e^-\to W^-\nu_{e}$ & Anomalous $W$ couplings & \cite{machado} \\
$\gamma\gamma\to WWWW$, $WWZZ$& Strong $WW$ scatt., 
quartic anomalous $W,Z$ coupl's & Insufficient $\sqrt{s}$ \\
\hline
$\gamma\gamma\to t\bar{t}$ & Anomalous top quark interactions &
  Low rate $\to$ use $\gamma g$ \\
$\gamma e^-\to \bar{t} b \nu_e$ & Anomalous $Wtb$ coupling & \cite{favereau} \\
\hline
$\gamma\gamma\to$ hadrons & Total $\gamma\gamma$ cross section & \cite{Godbole:2003hq} \\
$\gamma e^-\to e^- X$ and $\nu_{e}X$ & NC and CC structure functions
(pol.\ and unpol.) & ? \\ 
$\gamma g\to q\bar{q},\ c\bar{c}$ & Gluon distribution in the photon & ? \\
$\gamma\gamma\to J/\psi\, J/\psi $ & QCD Pomeron & \cite{kumar,nystrand} \\
\hline
\end{tabular}
\end{center}
\end{table*}
%%% End of Table 1 %%%
%
We have added an extra column to this table, pointing the reader
to perceived difficulties in adapting the respective production
channels to the LHC or, when applicable, to the relevant
contribution in these proceedings.

\section{Photon physics at hadron colliders}

With the realization of a photon collider option and the ILC itself
being now more uncertain than ever, the biggest advantage of using a
hadron collider for photoproduction experiments is that such
colliders (RHIC, the Tevatron and LHC) exist. The energies that can
be reached are in fact quite comparable: $\sqrt{s}_{\gamma\gamma}^
{\max} = 400$ GeV at a 500 GeV ILC vs.\ $\simeq 486$ GeV in OO
collisions at the LHC and even more in pp collisions. In addition,
hadron colliders offer the possibility to not only study
photon-photon, but also photon-proton or photon-ion collisions, in
particular at low $x$-values \cite{strikman}.

The disadvantages are quite obvious: Much less bremsstrahlung is
emitted by the heavier protons than by electrons, but this loss in
luminosity may be compensated by the large electrical charge $Z$ of
heavy ions. Elastic scattering of nucleons and nuclei is not only
induced by photons, but also by pomerons (also reggeons and possibly
odderons), and the two are not easily separated; one option is to
consider only electroweak final states, which do not couple to
pomerons. And then perturbative QED may no longer be applicable at
large $Z$, rendering a reliable luminosity calculation for heavy ion
beams difficult. In ultraperipheral heavy-ion collisions with
impact parameter $b$ larger than the sum of the colliding-ion radii,
radiation is emitted coherently by the whole nuclei, making them
vibrate, generating resonances and inducing large soft cross
sections and electron-positron pair production. The latter can
furthermore lead to electron capture and eventually beam loss
\cite{baur,guclu}. Fortunately, the calculational difficulties can be
overcome by using reactions like $\gamma\gamma\to l\bar{l}$ as a
luminosity monitor, as had already been proposed for the ILC.

In CMS this reaction has been studied and found to be of sufficient
quality once a $p_T$-cut of 3 (6) GeV for muon (electron) pairs has
been imposed and the lepton pair has been forced to have an azimuthal
angle $\Delta\phi$ larger than 2.9 (2.7) \cite{hollar}. Its
applicability may, however, be reduced to low-luminosity runs at the
LHC.

The Krakow-Paris collaboration estimates that the $p_T$-cut may even
be reduced to 0.2 (6) GeV or completely abandoned, provided that an
an additional detector allows to cut on $\Delta\phi>3.1$
\cite{krasny}.

In both cases, identifying photon interactions at hadron colliders
will probably require detection of the forward protons \cite{rouby,%
Piotrzkowski:2000rx}.
First, good knowledge of the proton transport in the beampipe is
required and can indeed be obtained with the HECTOR simulation tool,
validated by MAD-X. Second, the protons must be detected and their
energies measured at 420 and/or 220 meters from the primary vertex
with the FP420 and/or FP220 detectors. This allows then to
determine the energy of the exchanged photon in the range 20 ... 120
GeV with a precision of 1 ... 2 GeV and/or in the range 120 ... 900
GeV with a precision of 10 ... 12 GeV. The virtuality of the photon can
be derived from the $p_T$ of the proton, which can be measured in the range
0.1 ... 1 GeV with a precision of approximately 0.14 ... 0.77 GeV.

High-energy photon collisions at hadron colliders have been reviewed
in \cite{baurreview} and more recently for ultraperipheral heavy-ion collisions
in \cite{strikmanreview} and for proton-proton scattering at the LHC
in \cite{louvain}.

\subsection{RHIC}

\begin{figure}
 \includegraphics[width=\columnwidth]{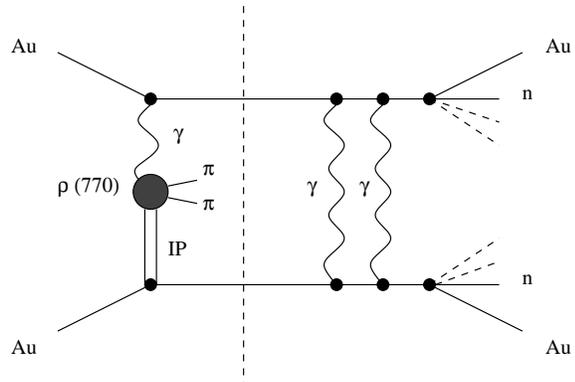}
 \caption{\label{fig:1}Typical Feynman diagram of photon-pomeron
 scattering in heavy-ion collisions, leading here to the production
 of a $\rho$ vector meson that decays subsequently into a pion
 pair. The nuclei (gold in this case) may be rescattered and/or
 excited, which may lead to additional forward-neutron production.}
\end{figure}

The Relativistic Heavy Ion Collider (RHIC) at BNL has started
operation in 2000 colliding nuclei with nucleon-nucleon center-of-mass energies
of up to 200 GeV, and photon interactions have indeed been
observed. A typical diagram is shown in Fig.\ \ref{fig:1}, where
%
%%% Begin  Figure 1 %%%
%%% End of Figure 1 %%%
%
a $\rho$ vector meson decaying to two pions is produced resonantly
in photon-pomeron collisions. The colliding gold nuclei are
sometimes excited and emit identifiable forward neutrons. These
events are characterized by a low multiplicity in the central
detector, typically only one primary vertex and two tracks with low
$p_T$.

The STAR collaboration has measured the rapidity distribution of the
produced pion pair and found agreement with theoretical calculations
based on (generalized) Vector-Meson Dominance (VMD) models, which
predict that the photon preferably fluctuates into hadronic states
with the same quantum numbers. Disagreement was found with a
calculation based on the color-dipole approach, but this may only be
a normalization problem of the photon flux. 
%Unfortunately, the flat measured rapidity
%distribution does not allow to distinguish the shapes, which are
%quite different in the two cases.
The distribution in the angle
$\phi$ between the $\rho$ decay and production planes has also been
measured and been found to be in agreement with $s$-channel helicity
conservation \cite{grube}. 

The production of 52 electron-positron pairs at rather low $p_T$ has
also been observed by STAR, making it clear that photon interactions
at hadron colliders are indeed happening. Both the invariant-mass
and $p_T$ distributions agree with full QED calculations taking into
account nuclear Coulomb excitation, e.g.\ of the giant dipole
resonance. The factorized approach based on the Equivalent-Photon
or Weizs\"acker-Williams Approximation (EPA or WWA) also describes
the data well with the exception of the very lowest $p_T$-bin
\cite{grube}.

The PHENIX collaboration have observed exclusive $J/\psi$
photoproduction, again in
gold-gold collisions at $\sqrt{s}_{NN}=200$ GeV, in the $e^+e^-$
decay channel at central rapidity. The measured cross section
d$\sigma/$d$y\,(y=0)=44\pm16$ (stat.) $\pm18$ (syst.) $\mu$b agrees
well with predictions based on coherent photon radiation and/or
quasi-elastic scattering, but one has to caution that so far only a
single data point at $y=0$ is available \cite{white}.

\subsection{Tevatron}

After a first run during the years 1992 to 1996, the (slightly)
higher hadronic center-of-mass energy of $\sqrt{s}=1.96$ TeV and, more
importantly, the higher luminosity and the considerably improved
detectors available since 2001 at Run II of the Tevatron have made
it possible to observe photon interactions there in three different
channels: $e^+e^-$, $\gamma\gamma$ and quarkonium production. The
CDF collaboration have managed to do so after accumulating 532
pb$^{-1}$ and 1.48 fb$^{-1}$ by using only their miniplug
calorimeter covering the range $|\eta|\in[3.5;5.5]$ and their beam
shower counters in the range $|\eta|\in [5.5;6]$, but no Roman pots
for proton/antiproton identification \cite{pinfold}.

16 events of exclusive $e^+e^-$ production with $E_T>4$ GeV have
been observed, where the inclusive and dissociation background was
only expected to be 1.9$\pm0.3$ events. The signal corresponds to a
total cross section of 1.6$^{+0.5}_{-0.3}$ (stat.) $\pm0.3$ (syst.)
pb and agrees nicely with the QED prediction of the LPAIR Monte
Carlo ($1.71\pm0.01$ pb) \cite{lpair}, as do the distributions in
invariant mass and azimuthal angle.

Three photon pairs have also been observed in the same data sample.
Since photoproduction of photons is a one-loop process, these events
come rather from double-pomeron exchange (or gluon-gluon scattering
plus initial-state rescattering), and one is even likely be due to
pion decay. Subtracting this event one obtains a total cross section
of 90$^{+120}_{-30}$ (stat.) $\pm16$ (syst.) fb. The theoretical
prediction lies a factor of three below \cite{khoze}. However, as
discussed above for diffractive photoproduction at HERA,
calculations of the rapidity-gap survival probability may still
have large theoretical uncertainties.

A preliminary analysis with the higher luminosity has lead to 334
events with muon pairs with clearly visible $J/\psi$ and $\psi'$
mass peaks and many candidate events pointing to intermediate
$\chi_c$ production. At higher masses, also the $\Upsilon$(1S) and
its 2S and possibly 3S excitations are visible in a sample of 145
events. The quarkonium analyses are, however, yet to be finalized
and published.

\subsection{LHC}

In contrast to electrons or positrons, for which the energy spectrum
of the radiated photons can be simply described by the equivalent
photon approximation for a charged pointlike particle \cite{fermi},
protons and nuclei have constituents, which make it necessary to
take in addition their charge distributions into account through
inelastic or elastic form factors.

(Valence) quarks have charges similar to the one of the proton
itself, so that inelastic exceed elastic photoproduction cross
sections for proton collisions. Form factors can be reliably
computed in this case in the plane-wave formalism.

The nuclear charge $Z$ can, on the other hand, be much larger than
one, so that elastic scattering of heavy ions can lead to strong
coherent radiation of photons with wavelength (or inverse
energy/transverse-momentum/virtuality) larger than the nuclear
radius $R\simeq1.2A^{1/3}$ (with $A$ being the nucleon number),
i.e.\ the nuclear structure is not resolved. The spectrum for
strong electromagnetic fields may be more reliably calculated in
the semiclassical approximation in impact parameter ($b$) space.
This makes it also possible to take into account absorptive
corrections from strong initial-state interactions, which are,
however, generally excluded by imposing $b>R_1+R_2$. By imposing
$b_{1,2}>R_{1,2}$, also strong interactions from produced hadrons
with the initial ions may be excluded.

In Tab.\ 2 we summarize the nucleon-nucleon center-of-mass energies
%
%%% Begin  Table 2 %%%
\begin{table*}
 \begin{center}
 \caption{Nucleon-nucleon center-of-mass energies and luminosities
 and the resulting maximal photon-nucleon and photon-photon
 center-of-mass energies for
 various colliding ion combinations at the LHC.}
 \begin{tabular}{lllll}
 \hline
 NN'  & $\sqrt{s}_{NN'}$/TeV & ${\cal L}_{NN'}$/mb$^{-1}$s$^{-1}$ &
 $\sqrt{s}_{\gamma N}$/GeV & $\sqrt{s}_{\gamma\gamma}$/GeV\\
 \hline
 \hline
 OO   & 7   & 160  & 1850 & 486 \\
 ArAr & 6.3 & 43   & 1430 & 322 \\
 PbPb & 5.5 & 0.42 & 950  & 162 \\
 \hline
 pO   & 9.9 & 10000& 2610 & 686 \\
 pAr  & 9.39& 5800 & 2130 & 480 \\
 pPb  & 8.8 & 420  & 1500 & 260 \\
 \hline
 pp   & 14  &$10^7$& 8390 & 4504 \\
 \hline
 \end{tabular}
 \end{center}
\end{table*}
%%% End of Table 2 %%%
%
and luminosities and the resulting maximal photon-nucleon and photon-photon
center-of-mass energies for various colliding ion combinations at
the LHC.

\subsubsection{ATLAS, CMS, and ALICE forward detectors}

During this workshop, the forward capabilities of three of the four main LHC
detectors were discussed: ATLAS, CMS, and ALICE. It is interesting to note that
most of the LHC collision energy in pp collisions will be deposited in the
rapidity range $6<|\eta|<8$, while the main ATLAS and CMS calorimeters only
cover the range $|\eta|<5$, ALICE even only the range $|\eta|<0.9$.
The forward detectors will therefore play an essential role in the
LHC physics program, not only for QCD studies at low-$x$ and in
diffraction, but also to identify high-energy photon collisions
in pp or heavy-ion collisions.

While the main ATLAS hadronic calorimeter allows already for the
selection of single or double rapidity-gap events, the identification of
elastically scattered protons requires additional forward detectors.
The LUminosity Cerenkov Integrating Detector (LUCID) is currently
being installed in the ATLAS cavern and will cover the range
$5.4<|\eta|<6.1$. As for the Zero Degree Calorimeter (ZDC), only a
simplified version will be (at least initially) installed and allow
to cover very large rapidities of $|\eta|>8.1$. The installation of
Roman pots to measure the Absolute Luminosity For ATLAS (ALFA) with
2 to 3\% accuracy is planned in mid-2009. They would cover the
range $10<|\eta|<14$ \cite{giacobbe,tasevsky}.

The CMS rapidity coverage will be extended by CASTOR to $5.2<|\eta|<
6.6$, thereby also enhancing its hermiticity, although funding has
so far only been made available for a detector on one side of the
interaction point to be installed in July 2008. Similarly to ATLAS,
a Zero Degree Calorimeter (ZDC) with rapidity coverage above $|\eta|
>8.4$ is already installed. The particularity of CMS is its symbiosis
with TOTEM, a relatively independent experiment hoping to measure
the total cross section and LHC luminosity with 1\% accuracy using
forward tracking detectors \cite{grothe}.

TOTEM and ATLAS (ALFA) Roman pots will be located at 220 m (FP220) covering
the proton longitudinal momentum fractions in the range 0.02 $< x_{1,2}<$ 0.2.
Both ATLAS and CMS have advanced plans to install high-precision
silicon tracking and
fast timing detectors at 420 m (FP420), which would cover the range
$0.002<x_{1,2}<0.02$ \cite{Albrow:2008pn}.
Since photon events have lower $x$-values than
pomerons, these detectors might prove indispensable to study photon
interactions at high luminosities. If approved, they could be
installed with either experiment in 2010 \cite{roeck}.

While the ALICE central hadronic calorimeter covers only the
rapidity range $|\eta|<0.9$, its particularity consists in a $p_T$
threshold which at 0.1 GeV is much lower
than those of ATLAS (0.5 GeV) and CMS (0.2 GeV). Additional forward detectors such
as the muon spectrometer ($2.5<|\eta|<4$) and particularly the
neutron Zero Degree Calorimeter should allow for energy vetoes and
thus the classification of single- or double-diffractive and
possibly photon events. However, no installation of Roman pots to
identify elastically scattered protons is planned at this point
\cite{schicker}.

\subsubsection{Quarkonium production}

One of the most interesting channels for high-energy photon
collisions at the LHC will be the production of vector mesons such
as heavy quarkonia. As their exclusive photoproduction proceeds
through double-gluon exchange from the heavy-ion target, this
channel will provide a very sensitive test of the nuclear gluon
density, which is essentially unknown below values of $x\simeq0.1$
\cite{d'Enterria:2007pk}.

\begin{figure}
 \includegraphics[width=\columnwidth]{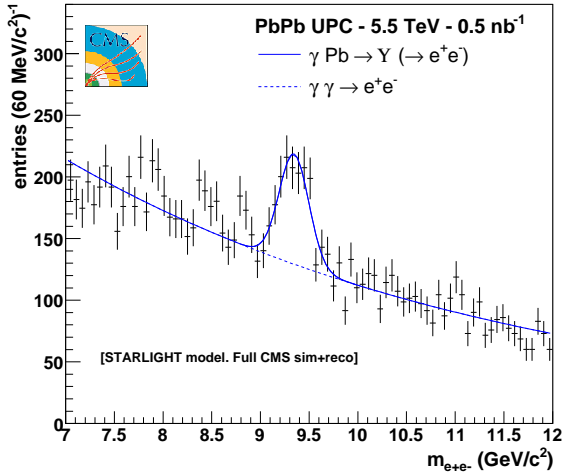}
 \caption{\label{fig:2}Simulated invariant-mass distribution for
 exclusive di-electron production through $\gamma\gamma$ scattering
 or the decay of photoproduced $\Upsilon$ mesons in PbPb collisions
 at $\sqrt{s} = 5.5$ TeV in CMS \cite{d'Enterria:2007pk,kumar}.
 A similar result is
 obtained for di-muon production.}
\end{figure}

The potential of the CMS detector to measure exclusively produced
$\Upsilon$(1S) mesons in ultraperipheral PbPb collisions is quite
promising with an expected detection rate of about 500 events for
an integrated luminosity of 0.5 nb$^{-1}$. This estimate is based
on a Starlight Monte Carlo simulation, giving a signal cross section
of 173 $\mu$b and background cross sections of 2.8 and 1.2 mb in the
electron and muon decay channels. The invariant-mass distribution of
the lepton pairs produced in PbPb collisions at $\sqrt{s}=5.5$ TeV
is shown in Fig.\ \ref{fig:2}, where the peak due
%
%%% Begin  Figure 2 %%%
%%% End of Figure 2 %%%
%
to the $\Upsilon$ resonance is clearly visible with a mass resolution
of about 150 MeV \cite{d'Enterria:2007pk,kumar}.

For ALICE one may expect the detection of about 400 $\Upsilon$ mesons,
which will be visible only in the electron channel. However, about 
$10^5$ $J/\psi$ mesons and even as many as $2\cdot 10^8$ $\rho$ mesons
will be produced and detected. For exclusively photoproduced
$J/\psi$ mesons with only 1.5h of design luminosity, the invariant-mass
distribution has been simulated for the electron decay channel.
 %is shown in Fig.\ \ref{fig:3}.
%
%%% Begin  Figure 3 %%%
%\begin{figure}
% Plot from CMS-AN06-107
% \caption{\label{fig:3}Simulated invariant-mass distribution for
% exclusive di-electron production through $\gamma\gamma$ scattering
% or the decay of photoproduced $J/\psi$ mesons in Pb-Pb collisions
% at $\sqrt{s} = 5.5$ TeV in ALICE \cite{nystrand}.}
%\end{figure}
%%% End of Figure 3 %%%
%
 %Again the mass peak due to the quarkonium resonance is very well
 %visible.
In pp collisions, where 1400 reconstructed $J/\psi$ mesons
are expected with 250h of ALICE luminosity, the signal will even be
much cleaner \cite{nystrand}.

The production mechanism for inclusively produced quarkonia is still
not fully understood. While NLO color-singlet calculations correctly
predict the $p_T$-spectrum of $J/\psi$ mesons in photoproduction at
HERA, LO predictions for hadroproduction at the Tevatron fail by more
than one order of magnitude and require the introduction of additional
color-octet contributions as predicted by NRQCD. $J/\psi$-production
in photon-photon collisions at LEP can then be correctly described
\cite{mihaila}, but this is unfortunately not the case for the
polarization of hadroproduced quarkonia. More experimental information
such as the one expected from the LHC may help to clear up the puzzle.
This is particularly important, as $J/\psi$ suppression remains to be one of
the key signatures of the quark-gluon plasma. The $p_T$-distributions
for inclusive $J/\psi$ and $\Upsilon$ photoproduction have been
evaluated for this workshop for pp as well as pPb and PbPb
collisions, using the Monte Carlo program MadOnia and taking into account
additional photon and gluon as well as open $c\bar{c}$ radiation
\cite{lansberg}.

\subsubsection{Anomalous gauge-boson and top-quark couplings}

A second promising possibility for high-energy photon collisions at the
LHC is the determination of anomalous vector-boson couplings, which can
either be parameterized in a process-specific way by form factors or
process-independently with effective Lagrangians after or before
electroweak symmetry breaking. In the latter case, the Lagrangian
can be constrained by imposing the equations of motion and lepton/baryon
number conservation, leaving only ten dimension-six operators with their
dimensionless couplings $h_i\sim{\cal O}(v^2/\Lambda^2)$ in the effective
Lagrangian, four of which are $CP$-violating. These may be determined
in $W$-boson pair production not only at the ILC, but also at the LHC,
although with less precision. When clean lepton final states and
optimal observables from fully differential cross sections are chosen,
the precision for the couplings not involving Higgs or $B$-bosons might be
considerably improved at the LHC over present bounds from LEP, SLD and
the Tevatron \cite{manteuffel}.

The triple ($WW\gamma$) \cite{kepka} and quartic ($WW\gamma\gamma$) \cite{pierzchala} gauge-boson couplings have been investigated with an
effective Lagrangian after electroweak symmetry breaking. With 30 fb$^{-1}$
of luminosity, the former may be determined with two to 15 times higher
precision than presently available from the Tevatron, while the latter
might even be improved by a factor of 10$^4$ over present LEP bounds with
an integrated luminosity of 10 fb$^{-1}$. However, these couplings are most
strongly enhanced at high invariant masses, where unitarity becomes
violated. Implementing a unitarity bound through an energy-dependent form
factor then creates unfortunately some model-dependence.

The unitarity problem arises also when trying to constrain the quartic
couplings in the pp-mode of LHC through $W$-boson fusion in the channel
$WW\to\gamma\gamma$. A considerable improvement over current LEP bounds
and a precision similar to the photon-photon case may be achieved
\cite{eboli}.

The production of single $W$-bosons has been measured at HERA by combining
the full Run-I and Run-II data sets with a total luminosity of about 1
fb$^{-1}$ and the H1 and ZEUS analyses. A small excess is observed in the
H1 data sample, which is, however, not significant in the combined
analysis. A search for anomalous single top-quark production in
flavor-changing neutral currents has also been performed, leading to the
current world's best limit on the magnetic coupling $\kappa_{tu\gamma}<0.14$
\cite{south}.

Such a coupling would lead to an LHC photoproduction cross section that
is about 100 times larger than the HERA cross section, so that there may
be much room for improvement. For low (high) luminosities of 1 (30)
fb$^{-1}$, one may expect the magnetic up-quark coupling limit to improve
to 0.044 (0.029), and a first limit on $\kappa_{tc\gamma}$ of about 0.077
(0.050) may be obtained. At the same time, the CKM-matrix element $V_{tb}$
can be measured, albeit only with a similar precision to the one already
achievable in regular pp collisions \cite{favereau}.

\subsubsection{Higgs-boson production}

While the resonant production of Higgs bosons is clearly one of the main
motivations for adding a photon collider option to a future ILC, the
situation looks much less promising for Higgs production in photon-photon
collisions at the LHC. For masses of 120 GeV, a few events per year may
still be expected in the $b\bar{b}$ decay channel, in particular in OO
and pp collisions, but clearly not enough to perform the precision studies
of quantum numbers and couplings that would be possible at an ILC.
For masses of 185 GeV the rate (now in the four-lepton channel) drops even
to less than one event per year \cite{sarycheva}.

The situation is slightly better for the associated photoproduction of
$W$ and Higgs bosons, which constitutes about 5\% of the total inclusive
rate.
With an integrated luminosity of 100 fb$^{-1}$, a
significance of up to 3$\sigma$ may be achieved for masses of about
170 GeV by combining different semileptonic channels. However, this is
clearly not enough to render $WH$ photoproduction a discovery channel
\cite{ovyn}.

The event topology of Higgs production in photon-photon collisions is
quite similar to the now well-known weak-boson fusion channel, with the
important difference that rates are much higher here and two additional
forward jets are available for event selection. A central jet veto
allows to suppress most of the QCD background, so that a significance
of 5$\sigma$ can already be achieved with 30 fb$^{-1}$ for masses between
110 and 140 GeV in the $\tau\tau$ decay channel; for larger masses, the
$W$-boson decay channel may be used. With 200 fb$^{-1}$, the partial
widths and couplings
can be measured with 10-30\% and 5-15\% errors, respectively
\cite{zeppenfeld}.

\subsubsection{Slepton production}

In $R$-parity conserving supersymmetry (SUSY), sleptons decay into
leptons and a neutralino, which is often the lightest SUSY particle
(LSP) and thus escapes undetected. In inelastic hadron collisions,
these events are selected by triggering on missing transverse energy,
as the longitudinal-momentum balance from the colliding partons is
unknown. In photon-photon collisions, however, the forward protons
can be detected and their energy measured, thus providing additional
information on the total center-of-mass energy and the longitudinal
momenta of two charged decay leptons. This allows for improved
rejection of the Drell-Yan, $W$- and $\tau$-decay backgrounds and
better extraction of the slepton mass and slepton-neutralino mass
difference through kinematic edges \cite{gronberg}.

A detailed study of slepton-pair production has been performed for
the LM1 benchmark point, where a common scalar mass of $m_0=60$ GeV
induces light sleptons with masses between 100 and 200 GeV and thus
large signal cross sections of about 2.2 fb, which is only slightly
reduced by acceptance cuts to 0.7 fb \cite{schul}.
Additional background
suppression was achieved here by requiring the decay leptons to share
the same flavor and the events to be coplanar. A $5\sigma$ discovery
might then be achieved with only 25 fb$^{-1}$ of luminosity for
selectrons and muons, while for staus at least 100 fb$^{-1}$ would
be needed. The mass resolution should be of the order of a few GeV,
which would be very similar to the mass resolution achievable in
photon-photon collisions at the ILC \cite{berge}.

The production of light charginos has also been investigated in the
same study, since their decays into $W$-bosons and neutralinos may lead
to a similar signal, provided that decays into intermediate sleptons
are forbidden. This would be the case for heavy sleptons as predicted
by the LM9 benchmark point with a heavy common scalar mass of 1450
GeV. The important difference is that in this case the decay leptons
can also be of different flavor and acoplanar, so that background
suppression is less efficient and at least 100 fb$^{-1}$ of luminosity
would be needed for a $5\sigma$ discovery.
%
%%% Begin  Figure 3 %%%
\begin{figure}[h]
 \includegraphics[width=\columnwidth]{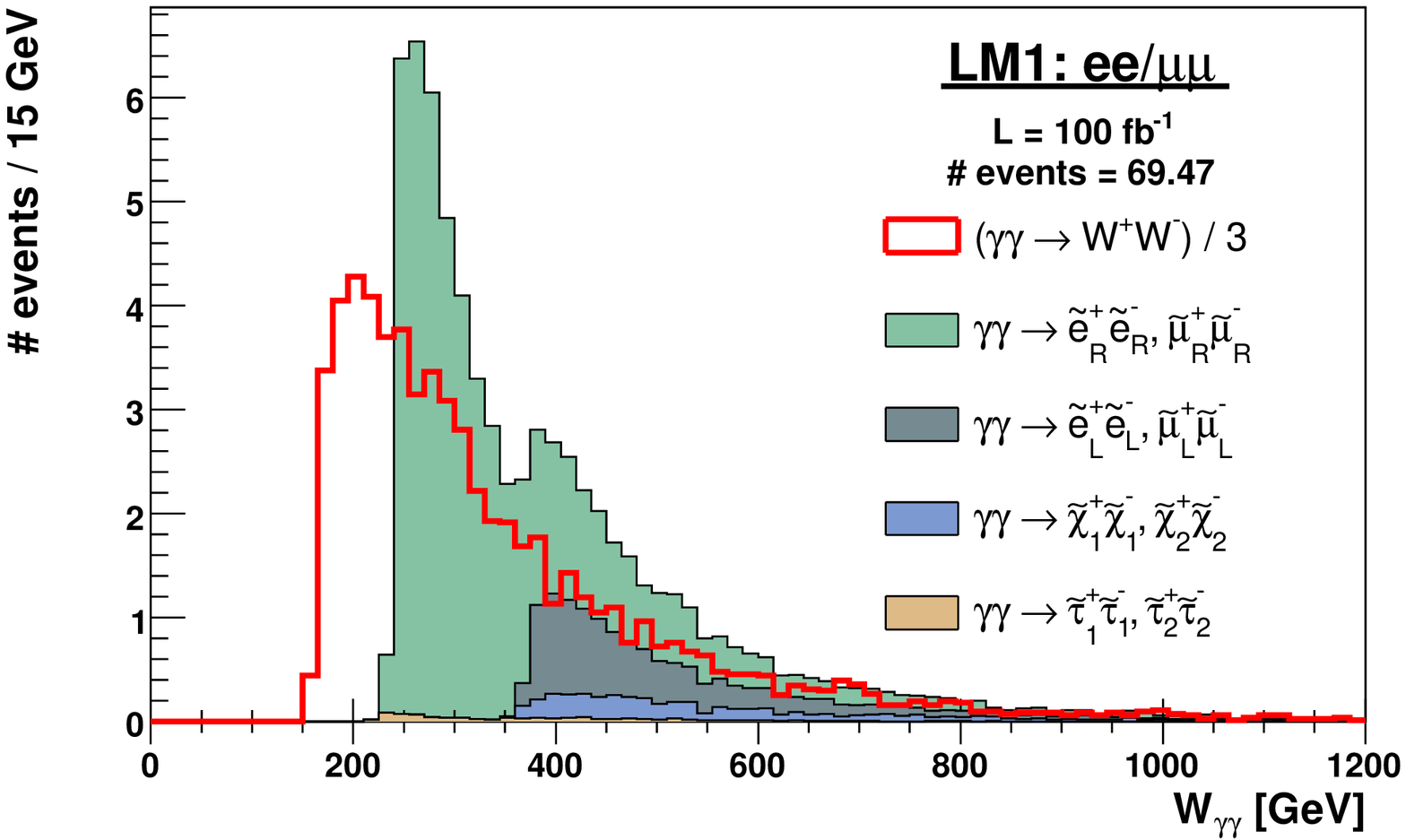}
 \includegraphics[width=\columnwidth]{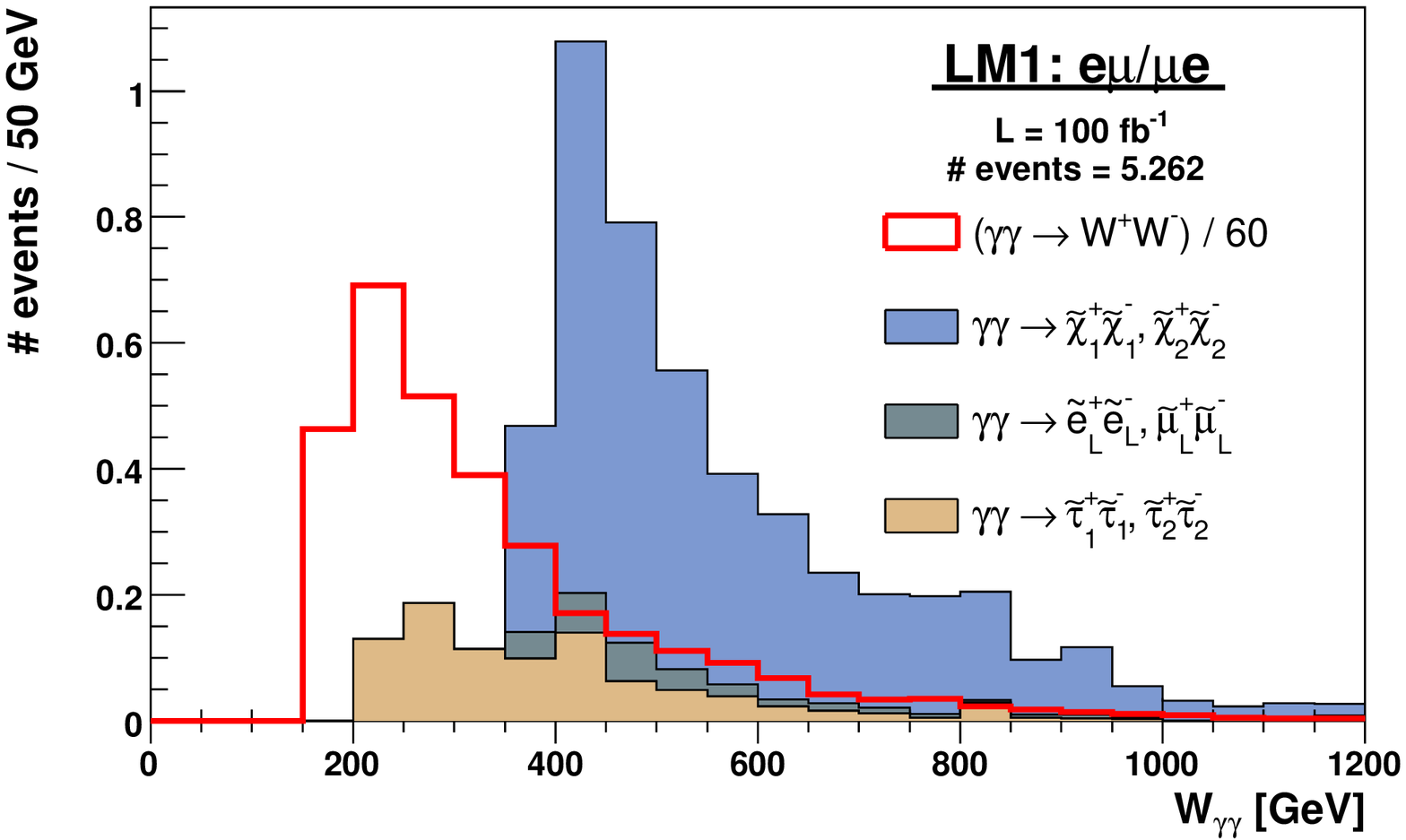}
 \caption{\label{fig:3}Photon-photon invariant mass for benchmark point LM1 with
 $\int\mathcal{L}dt$ = 100 fb$^{-1}$. Cumulative distributions for signal with two
 detected leptons ($p_T > 3$ GeV, $|\eta| <$ 2.5), two detected protons, with
 same (top) or different flavor (bottom). The $WW$ background has been
 down-scaled by the quoted factor \cite{schul}.}
\end{figure}
%%% End of Figure 3 %%%
%
In Fig.\ \ref{fig:3}, various SUSY channels are compared with the (rescaled)
background from $WW$ production \cite{schul}.

\section{Outlook}

After the shutdowns of LEP, HERA, and soon the Tevatron and in the
absence of an ILC, the LHC will provide an almost unique environment
to study high-energy photon collisions, rivaled only by the
continuing RHIC program. Event rates at the LHC will in fact be
dominated by forward scattering and include many elastic events and
low-level nuclear excitation. Proton (and neutron) identification will be
crucial to exploit these events, and as we have seen a number of
detectors are either already being installed or planned to match
this purpose in the ALICE, ATLAS and CMS experiments.

Traditionally, forward scattering has been the domain of diffractive
or low-$x$ QCD studies, and elastic vector-meson production is indeed
one of the promising channels which may allow for a better
determination of the proton's low-$x$ gluon density, pomeron slope,
and maybe discovery of the elusive odderon. However, also the poorly
understood inclusive quarkonium production mechanism may be elucidated,
and open jet and light- or heavy-quark production will provide unique
channels to determine the badly constrained
nuclear parton densities. This domain should
clearly be investigated further and in particular by adapting the
existing NLO codes from the HERA and LEP analyses to the LHC environment.

A precision determination of the top-quark charge might also be possible,
but here the potential of photon-photon and photon-proton scattering
has to be compared with the one of associated $t\bar{t}\gamma$
production in inelastic hadron collisions \cite{Baur:2001si}.

The sensitivity for anomalous couplings of vector bosons certainly looks
also very promising, as we have seen in several studies presented at
this workshop. However, these studies are still lacking a common
theoretical framework. They should be based on the same effective
Lagrangian, implement the unitarity bound in the same way, and agree on
a common set of old limits to be improved, so that photon-photon
scattering can be reliably compared with weak-boson fusion.

Unfortunately, the production of Higgs bosons seems pretty hopeless in $\gamma\gamma$ and difficult in $\gamma p$ scattering. SUSY Higgs bosons
and in particular charged Higgs bosons might be more promising, but have
not been discussed here. More work along these lines is clearly needed.

As we have seen, sleptons and charginos are certainly one of the promising
channels in the realm of physics beyond the Standard Model.
Apart from sleptons and gauginos, the SUSY spectrum includes also strongly
interacting squarks and gluinos. They therefore receive contributions not
only from photon, but also pomeron exchange, and it might be interesting,
although challenging, to compare and disentangle both contributions.
Extended SUSY models, such as those containing $R$-parity violation or extended
scalar sectors as in the NMSSM, and other models, such as extra-dimensional
or little-Higgs models, would certainly also be worth a close look.

In conclusion, this first workshop on ``High-energy photon collisions at the
LHC'' has proven that there is a large potential for interesting physics
studies and opened many perspectives for improvement of existing and
undertaking of a wide variety of future studies. The community is therefore
looking forward to many years of stimulating scientific discussion in this
field.

\section*{Acknowledgments}

The author thanks D.\ d'Enterria and K.\ Piotrzkowski for a careful reading of
and many useful comments on the manuscript as well as for the pleasure to
co-organize this exciting workshop with them.

\end{document}